\documentclass{article}
\usepackage{arxiv}
\usepackage{indentfirst}
\usepackage[utf8]{inputenc} % allow utf-8 input
\usepackage[T1]{fontenc}    % use 8-bit T1 fonts
\usepackage{hyperref}       % hyperlinks
\hypersetup{
    colorlinks=true,
    linkcolor=blue,
    filecolor=magenta,      
    urlcolor=cyan,
}
\usepackage{url}            % simple URL typesetting
\usepackage{booktabs}       % professional-quality tables
\usepackage{amsfonts}       % blackboard math symbols
\usepackage{nicefrac}       % compact symbols for 1/2, etc.
\usepackage{microtype}      % microtypography
\usepackage{graphicx}
\usepackage[ruled,vlined]{algorithm2e}
\usepackage{amsmath}
\usepackage{float} 
\usepackage{xcolor,colortbl}
\usepackage{changepage}  
\usepackage[square,numbers]{natbib}
\begin{document}

\title{Why Temporal Persistence of Biometric Features is so Valuable for Classification Performance}

\author{
Lee Friedman\\
Department of Computer Science\\
Texas State University\\
San Marcos, Texas, 78666, USA\\
\texttt{l\_f96@txstate.edu}
\and
Hal Stern\\
Department of Statistics\\
University of California - Irvine\\
Irvine, California, 92697, USA\\
\texttt{sternh@uci.edu}\\
\and
Larry R Price\\
Methodology, Measurement\\
and Statistical Analysis\\
Texas State University\\
San Marcos, Texas, 78666, USA\\
\texttt{lprice@txstate.edu}\\
\and
Oleg V Komogortsev\\
Department of Computer Science\\
Texas State University\\
San Marcos, Texas, 78666, USA\\
\texttt{ok11@txstate.edu}
}
\maketitle

\begin{abstract}
 It is generally accepted that relatively more permanent (i.e., more temporally persistent) traits are more valuable for biometric performance than less permanent traits. Although this finding is intuitive, there is no current work identifying exactly where in the biometric analysis temporal persistence makes a difference. In this paper, we answer this question. In a recent report, we introduced the intraclass correlation coefficient (ICC) as an index of temporal persistence for such features. In that report, we also showed that choosing only the most temporally persistent features yielded superior performance in 12 of 14 datasets. Motivated by those empirical results, we present a novel approach using synthetic features to study which aspects of a biometric identification study are influenced by the temporal persistence of features. What we show is that using more temporally persistent features produces effects on the similarity score distributions that explain why this quality is so key to biometric performance. The results identified with the synthetic data are largely reinforced by an analysis of two datasets, one based on eye-movements and one based on gait. There was one difference between the synthetic and real data:  In real data, features are intercorrelated, with the level of intercorrelation increasing with increasing ICC. This increasedhttps://www.overleaf.com/project/5e2b14694c5dc600017292e6 intercorrelation in real data was associated with an increase in the spread of the impostor similarity score distributions. Removing these intercorrelations for real datasets with a decorrelation step produced results which were very similar to that obtained with synthetic features. 
\end{abstract}

\section{Introduction}
It is generally accepted that relatively more permanent traits are more valuable for biometric performance than less permanent traits \citep{RN1603,RN1621,RN1623,RN1567,RN1619}. Since no human trait is permanent and true permanence is not required for biometric performance, we prefer the more precise term, ``temporally persistent''. Although from a conceptual point of view, this is quite intuitive, there is currently no understanding about exactly where in the biometric analysis temporal persistence makes a difference. This is the question addressed in this report. 

In a recent report \citep{RN1513}, we introduced the intraclass correlation coefficient (ICC) as an index of temporal persistence (e.g., stability, permanence) of single biometric features. In that report and the present report, we are exclusively dealing with features which are normally distributed or can be transformed into a normal form. The ICC can only be calculated if each subject is tested on 2 or more occasions.  For a biometric system, with multiple features available for selection, the ICC can be used to measure the relative stability of each feature. In the report, we also showed that choosing only the most temporally persistent features yielded superior performance in 12 of 14 datasets (p = 0. 0065, one-tailed).  Thirteen of the 14 datasets in that paper were real datasets from a number of different biometric modalities, including oculomotor, face, gait and brain structure. In general, then, for those datasets, prescreening potential biometric features, and choosing only highly reliable features yielded better performance than choosing lower ICC features or than choosing all features combined.    For eye movement-driven biometrics, the use of reliable features, as measured by ICC, allowed to us achieve an equal error rate (EER) 2.0\%, which was not possible before.  In that report, we did make some attempt to answer the question addressed by this manuscript. We present this report to more fully, accurately and precisely answer the question posed by this manuscript.

In the service of this goal, we present here a method for creating synthetic datasets with a number of properties that are helpful for studying biometric performance. Because the data are synthetic, we are able to control the degree of temporal persistence of the features while also ensuring that features are approximately independent of each other and thus provide unique pieces of information for biometric verification. We think that having unique pieces of information will allow us to address several theoretical notions relevant to biometric analysis in this and subsequent studies. Section reviews the relevant literature. In Section, we present our method to create synthetic datasets and show the theoretical relationship between temporal persistence and the distributions of similarity scores that are used in biometric systems (Experiment 1). Sections and explore two biometric datasets, one based on eye-movement features (Experiment 2), and the other based on gait-related features (Experiment 3). In Section, we identify one aspect of the synthetic data that differs from the real data and investigate the impact of this difference on biometric performance. Section provides a closing discussion.  

\section{Prior work on permanence in biometrics}
\label{sec:prior}
There are many reports which assess either template aging or permanence that operate at the level of a complete biometric system rather than at the level of single features \citep{RN1625,RN911,RN1411,Scheidat2012ShortTT,RN1627,RN1388,RN1450,RN1393,RN1451,RN1453} which is our interest here. For example, one recent paper \cite{RN1623} proposes a new method for measuring biometric permanence. Although these authors state that their method provides estimates of the permanence of biometric features, their measure of permanence is actually at the level of a complete biometric system rather than at the level of individual features. This is also true for the report on the permanence of ECG biometrics \cite{RN1619}, where permanence is assessed by looking at plots of EER over various time intervals. 

Jain \cite{1262027} discusses the importance of the permanence of biometric features but does not provide a method for assessing the permanence of individual features. In our prior paper \cite{RN1513}, we introduced the ICC for the assessment of the persistence of individual biometric features. Another recent paper \cite{RN1621} creates indices of permanence for brain waves. This method is specific for this modality, or at least for features which emerge from time-series (perhaps ECG, for example). 

A concept related to temporal persistence is that of reliable bits in binary or quantized biometric features, as it tries to identify those bits in biometric features that show minimal intra-subject variation. We may assume that these  bits will also be temporally persistent as long as the features themselves do not change with time. The idea of identifying reliable bits was first introduced in \cite{Tuyls} for binarized features derived from fingerprint templates. Here only features were binarized (quantized to 1 bit) that had a certain distance to the binarization threshold. In \cite{Chen1,Chen2}  more advanced methods to extract reliable bits from quantized features were proposed, in which the number of quantization levels per feature was chosen such that recognition performance is optimised, given a maximum total number of bits to encode the features. The optimisation assumes Gaussian probability density functions (PDFs) for the features, and turns out to work well in practice. In \cite{Tuyls,Chen1,Chen2}, additional information has to be added to the biometric template that indicates how features are quantized. 

In terms of accuracy we expect that the use of reliable bits as presented in \cite{Tuyls} in a metric for temporally persistence could result in at most similar performance as that of ICC, because the features are coarsely quantized to 1 bit. The reliable bits according to \cite{Chen1, Chen2} might have similar performance to the ICC as they are the result of a finer quantization. Further, empirical, research is needed to confirm these expectations. However, the computation of  reliable bits according to \cite{Chen1, Chen2} requires a constraint optimisation that is more computationally demanding than the computation of ICC.

\section{Creation and Analysis of Synthetic datasets}
\label{sec:creation}

\subsection{Creation of Synthetic Data}

Recall that the intraclass correlation coefficient (ICC) is a measure of the correlation expected for repeated measurements of the same feature on different occasions. Unlike the Pearson r correlation coefficient, which is typically applied as an interclass measure of relative agreement (i.e., two series can be correlated even if they differ substantially in level and spread), the ICC is an intraclass measure of absolute agreement~\cite{RN1513}. Measures from the same set of subjects at two different times are intraclass measurements (same metric and variance). ICC ranges from 0.0 to 1.0 with the latter corresponding to perfect temporal persistence. Our goal is to create synthetic features with a specified target ICC (denoted $ICC_{Target}$). Let $X_{ijs}$ denote the measurement of feature $j$ ($j = 1, \dotsc, K$) on session (occasion) $s$ ($s = 1, \dotsc, S$) for individual $i$ ($i = 1, \dotsc, N$). Although the ICC can be calculated based on many sessions, in our experience, biometric assessment is typically performed comparing only two points in time. Therefore, henceforth we will set $S = 2$. We generate normally distributed features such that the theoretical intraclass correlation of repeated measurements of the same feature on the same subject is $ICC_{Target}$ while the theoretical correlation of measurements of different features on the same individual and the theoretical correlation of measurements from different individuals are zero. In practice when data are simulated there are small variations in the empirical ICCs and there are small intercorrelations between features (and individuals) due to chance. 

The algorithm that we use is described briefly here and spelled out in Algorithm~\ref{alg:synthetic}. The starting point is to populate the full set of session one measurements $X_{ij1}$ with random draws from a standard normal distribution (mean zero and variance one). Then the measurements for the second session are set equal to the value of the given feature from the first session,  $X_{ij2} = X_{ij1}$ for ($i = 1, \dotsc, N$, $j=1,\ldots,K$
). At this point both sessions have the same data and each feature has ICC equal to 1.0 (perfect persistence). We obtain the desired ICC by adding a draw from a normal distribution with $\text{mean} = 0$ and $\text{variance} = (1 - ICC_{Target}) / ICC_{Target}$ to each of the measurements. At the end we apply a z-score transform to each feature (with all sessions concatenated together) so that they all have mean 0 and standard deviation one. It can be shown that the resulting measurements have the desired ICC (up to simulation noise). R code to create such synthetic datasets and to assess the resulting ICC is presented in Appendix I.

\begin{algorithm}
\caption{Creating Synthetic Features}
\SetKwInOut{Input}{Input}\SetKwInOut{Output}{Output}
\Input{$N$ (subjects), $K$ (features), $ICC_{Target}$}
\Output{3-dimensional ($N \times K \times 2$) feature matrix $X_{ijs}$ with desired correlation structure}
\hrule
\smallskip

for $j = 1, \dotsc K$\\
\quad    for $i = 1,\dotsc N$\\
\quad \quad Set $X_{ij1} = Z$ where $Z$ is a random standard normal deviate.\\
\quad \quad Set $X_{ij2} = X_{ij1}$ \\
\hrule
\smallskip
for $j = 1, \dotsc K$\\
\quad    for $i = 1,\dotsc N$\\
\quad \quad for $s = 1,2$\\
\quad \quad \quad Set $X_{ijs} = X_{ijs} + W$; where $W$ is a random normal deviate with mean = 0 and \\
\quad \quad \quad standard deviation = $\sqrt{(1 - ICC_{Target}) / ICC_{Target}}$\\
\hrule
\smallskip
For each feature $j$, treat $X_{ijs}$ as a single vector of length $N \cdot S$ and apply a z-score 
transform \\to ensure mean = 0 and standard deviation = 1
\label{alg:synthetic}
\end{algorithm}

Using this method, we can create features which are normally distributed, that have specified ICCs, with as many subjects and sessions as we desire. These features all have mean~=~0 and SD~=~1. These features are generally independent, but there are some small intercorrelations between features due to chance. To illustrate the approach, we generated data for 10000~subjects, 1000~features and 2~occasions with $ICC_{Target} = 0.7$. Figure 1(A) shows a histogram of the resulting empirical ICCs. Figure 1(B) shows a histogram of the resulting inter-feature correlations.

\begin{figure}
    \centering
    \includegraphics[trim={0.5em 0 1.5em 0}, clip, width=0.85\linewidth]{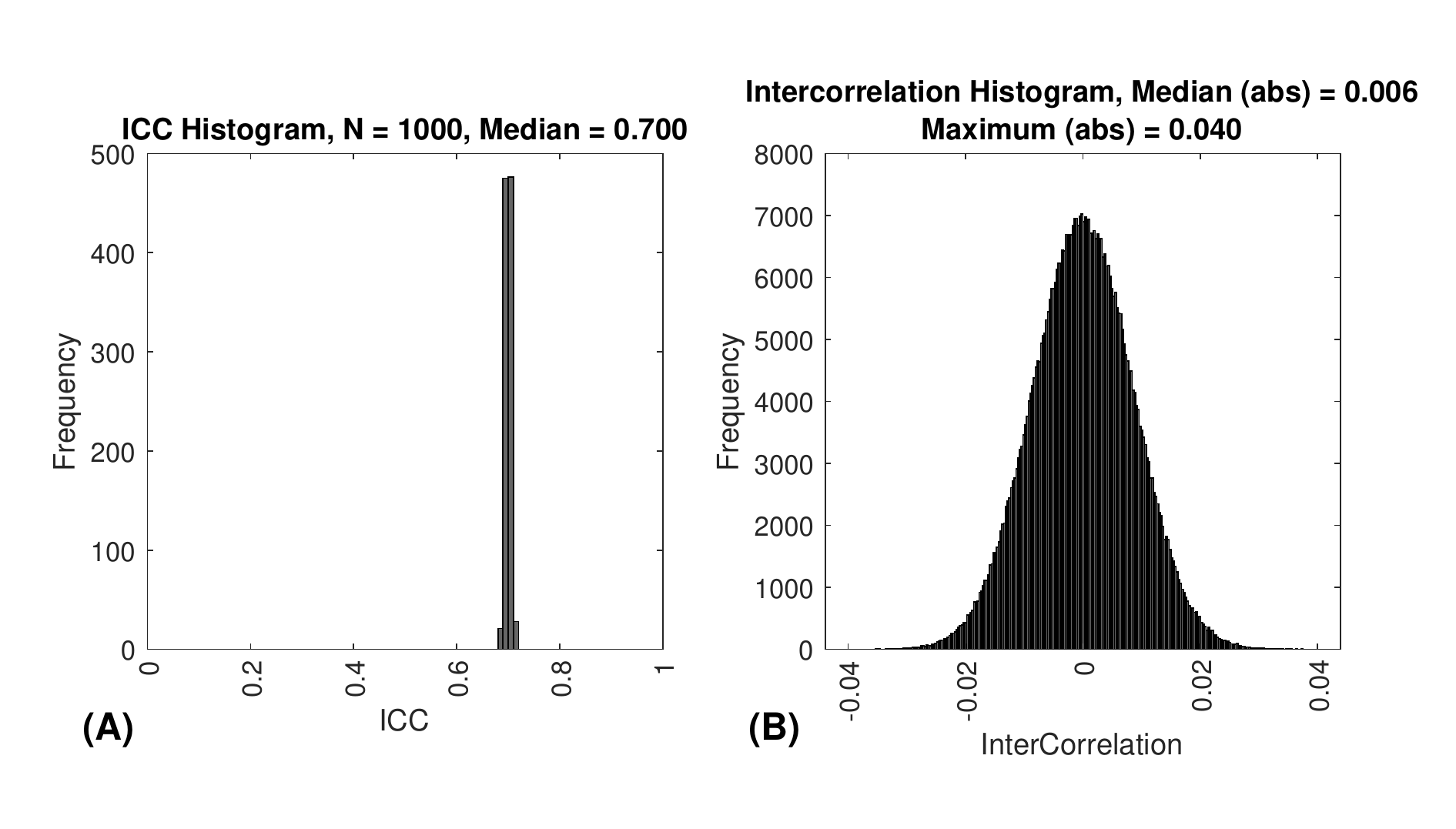}
\caption{(A)~Frequency histogram of ICCs for 1,000~features with an $ICC_{Target} = 0.7$. This is from a synthetic dataset with 10,000~subjects. (B)~Frequency histogram of correlations between 1,000~features for 10,000~subjects, two sessions, with an $ICC_{Target} = 0.7$. Note that the median and maximum are of the absolute value of the correlations.}
\end{figure}

\subsection{Creation of Sets of Features with Varying Degrees of Persistence}
To study the relationship of temporal persistence and biometric performance we generate a series of synthetic data sets with varying ICCs. To be specific we create 10~different datasets with each dataset consisting of 50~features and the features in each dataset having ICC values that vary over a small interval (e.g. 0.6~to~0.7). We denote the datasets as ``Bands'' to indicate that they cover different bands of the range of possible ICCs. Band~0 consists of 50~features simulated to have ICCs between 0.0~and~0.1 (with 5~features generated using Algorithm~\ref{alg:synthetic} with $ICC_{Target} = 0.005$, 5~features generated with $ICC_{Target} = 0.015$, \dots, 5~features generated with $ICC_{Target} = 0.095$). Band~1 has ICCs between 0.1~and~0.2 (again evenly spread out across that range), and so on through Band~9 which has ICCs between 0.9~and~1.0.

\subsection{Biometric Performance Assessment}
For each band (dataset), distance scores were calculated using only 20~randomly chosen features from the full set of 50~features in each band. We chose this number empirically based on the range of EER values produced across the bands.  We employed the cosine distance metric, since we have shown in an earlier (unpublished) report that the best biometric performance is produced with this choice \href{https://www.doi.org/10.13140/RG.2.2.17510.06727}{(Link to unpublished report)}. The resulting distance measures were scaled to go from 0~to~1 and then they were reflected ($1 - \text{distance}$) to compute similarity scores. A ``genuine'' distribution of similarity scores was constructed from the similarity scores for each subject and his/her self. All other similarity scores were considered impostors. These data were submitted to a ROC analysis and the EER was computed.

\subsection{Plotting Similarity Score Distributions for each Band}
After each ROC analysis for each band, we displayed the similarity score distributions, as in Figure 2. From inspection of these it became clear that the distributions of imposter similarity scores were not changing as we moved from Band~0 to Band~9, but that there were marked changes in the distributions of genuine similarity scores. The medians of the genuine distributions increase from Band~0 to Band~9, and the interquartile ranges (IQR) of the genuine distributions decrease from Band~0 to Band~9. These patterns are clearly shown in Figure 2 and Figure 3. 

\begin{figure}
    \includegraphics[trim={0 0 2em 0}, clip, width=0.95\linewidth]{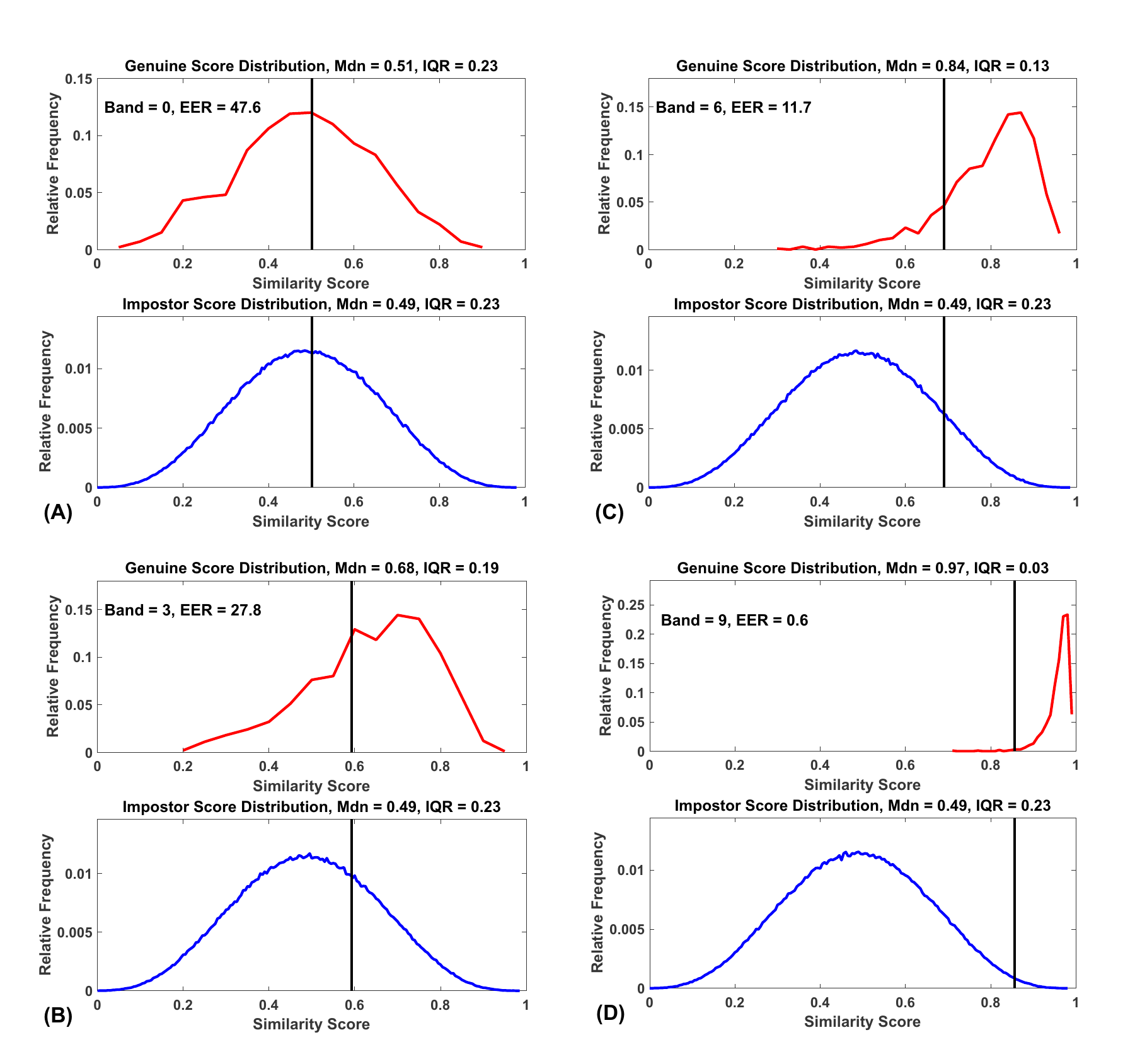}
    \caption{Similarity score distributions for Bands~0, 3, 6 and 9. (A)~Distributions for Band~0 (ICC from 0.0~to~0.1). (B)~Distributions for Band~3 (ICC from 0.3~to~0.4). (C)~Distributions for Band~6 (ICC from 0.6~to~0.7). (D)~Distributions for Band~9 (ICC from 0.9~to~1.0). Frequency is expressed as a proportion of all values. Mdn = median, IQR = interquartile range. Vertical black lines represent the threshold at which EER was achieved.}
\end{figure}

\begin{figure}
    \includegraphics[width=0.60\textwidth]{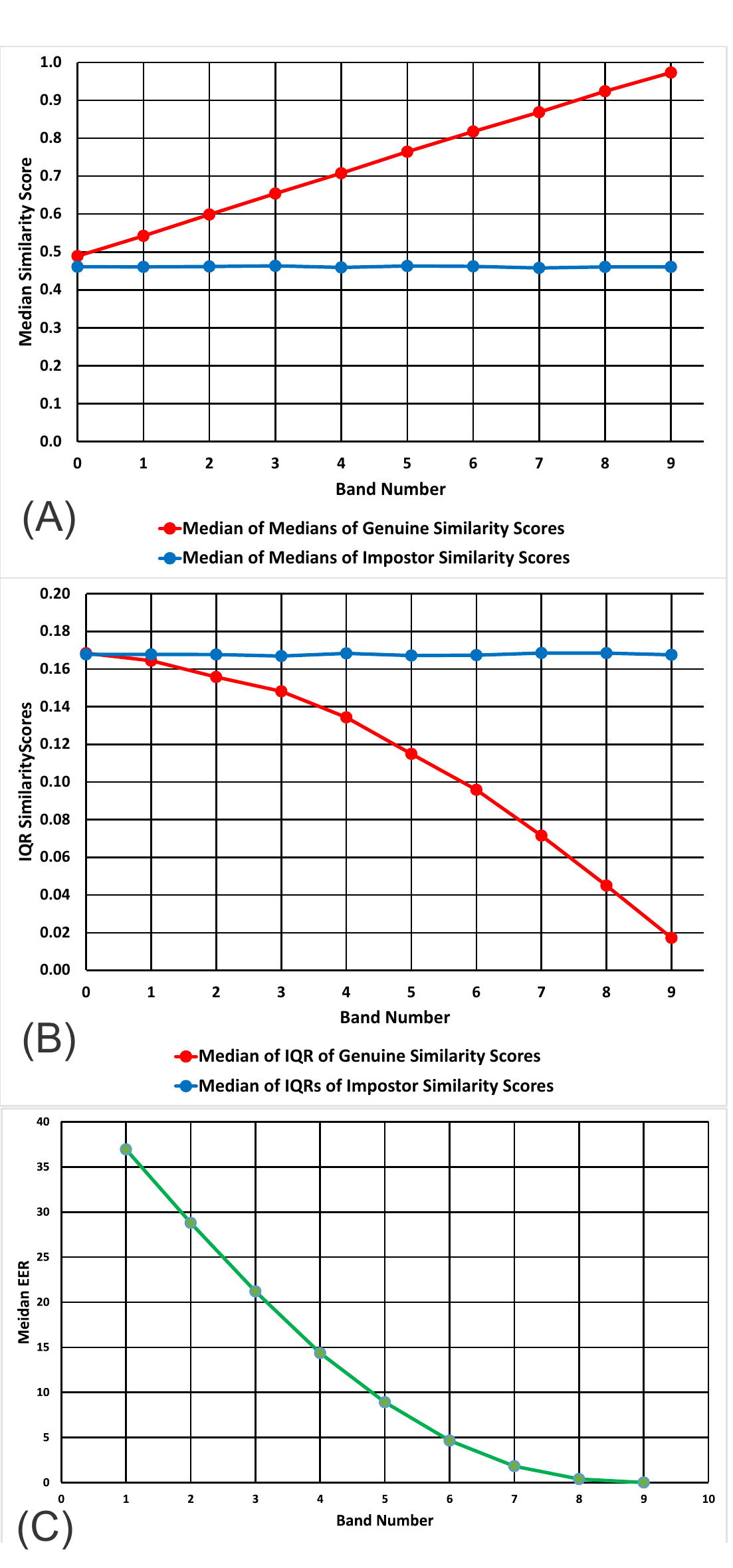}
    \caption{Medians and IQRs of genuine and impostor similarity scores and EERs across bands. (A)~Medians for synthetic data. Each point is the median of 10~median estimates. (B)~IQRs for synthetic data. Each point is the median of 10~IQR estimates. (C) Median EERs. Ranges for the values in this table can be found in Appendix Tables 1, 2 and 3.}
\end{figure}

\subsection{Discussion of Results}
The main findings of Experiment~1 are that using synthetic features with higher temporal persistence for biometric analysis produces lower EER values, i.e., improved biometric performance. These improvements are due to an increased median and a decreased IQR of the genuine similarity score distributions with no change in the impostor distributions. These results explain why features with increased temporal persistence produce better biometric results.

\section{Evaluation of the SBA Dataset}
\label{sec:eval-sba}

\subsection{Description of the SBA Dataset}
The real eye movement data we employed for this study came from 298~subjects recorded on two sessions. For more details, see~\citep{RN1513}. Subjects in the original study viewed 7~different tracking tasks. Only the text-reading task is relevant for the current report. Each subject was asked to read, silently, an identical pair of quatrains from the famous nonsense poem, ``Hunting for a Snark'', written by Lewis Carroll (written from 1874-1876). 	The EyeLink~1000 (SR Research Ltd., Kanata, ON, Canada), a video-oculography system which employs detection of both the pupil and the corneal reflection to determine gaze position, was used to record eye movements. It records both vertical and horizontal eye movements. In the present study, only left eye movements were collected. For 298~subjects, we have a mean spatial accuracy of 0.50~degrees of visual angle (SD~=~0.17, min~=~0.20, max~=~1.06). For further specifications, see the SR-Research website (\color{blue}\href{SR-Research Website} {https://www.sr-research.com/}\color{black}). The sampling rate for our data was 1000~Hz. Custom software transformed the raw records into gaze position data, in visual angle units, using the calibration data collected at the start of each task. The Stampe heuristic spike removal algorithm was employed~\cite{RN696}. In addition, blinks were detected and removed from the data by the EyeLink~1000. The eye movements were analyzed off-line. On each subject visit, subjects were studied twice (Sessions~1 and~2), approximately 20~min apart. They were given 60~seconds to read the poem passage. Session~1 to Session~2 (task-to-task) time intervals ranged from 13~min to 42~min (mean~=~19.5; SD~=~4.2). For eye movement classification, we employed the MNH~method described in~\citep{RN1647}. It identifies fixation periods, saccades, and post-saccadic oscillations~(PSOs) as well as periods of artifact and noise. Other portions of the recordings were left unclassified. For details regarding feature extraction, see~\citet{RN1516}.

\subsection{Biometric Assessment of SBA Dataset}
The ICCs for the SBA~dataset were available from~\cite{RN1513}. Our goal was to divide up the data into ``temporal persistence'' bands with equal numbers of features in each band. We found that if we created band limits based on ICCs from 0.0~to~0.9 in steps of 0.1, we would have at least 19~features per band.

For each dataset, standard ROC analyses were performed and similarity score distribution characteristics (median and IQR) were saved and plotted as a function of band number.

\subsection{Results with SBA Dataset}
In Figure 4(A), we present the medians for the genuine and impostor distributions, for bands~0 (ICCs: 0.0~to~0.1) to 8 (ICCS: 0.8~to~0.9) for the real SBA dataset. In Figure 4(B), we present the interquartile ranges for the genuine and impostor distributions, for bands~0 to~8 for the real SBA dataset.

\begin{figure}
    \centering
    \includegraphics[width=0.65\textwidth]{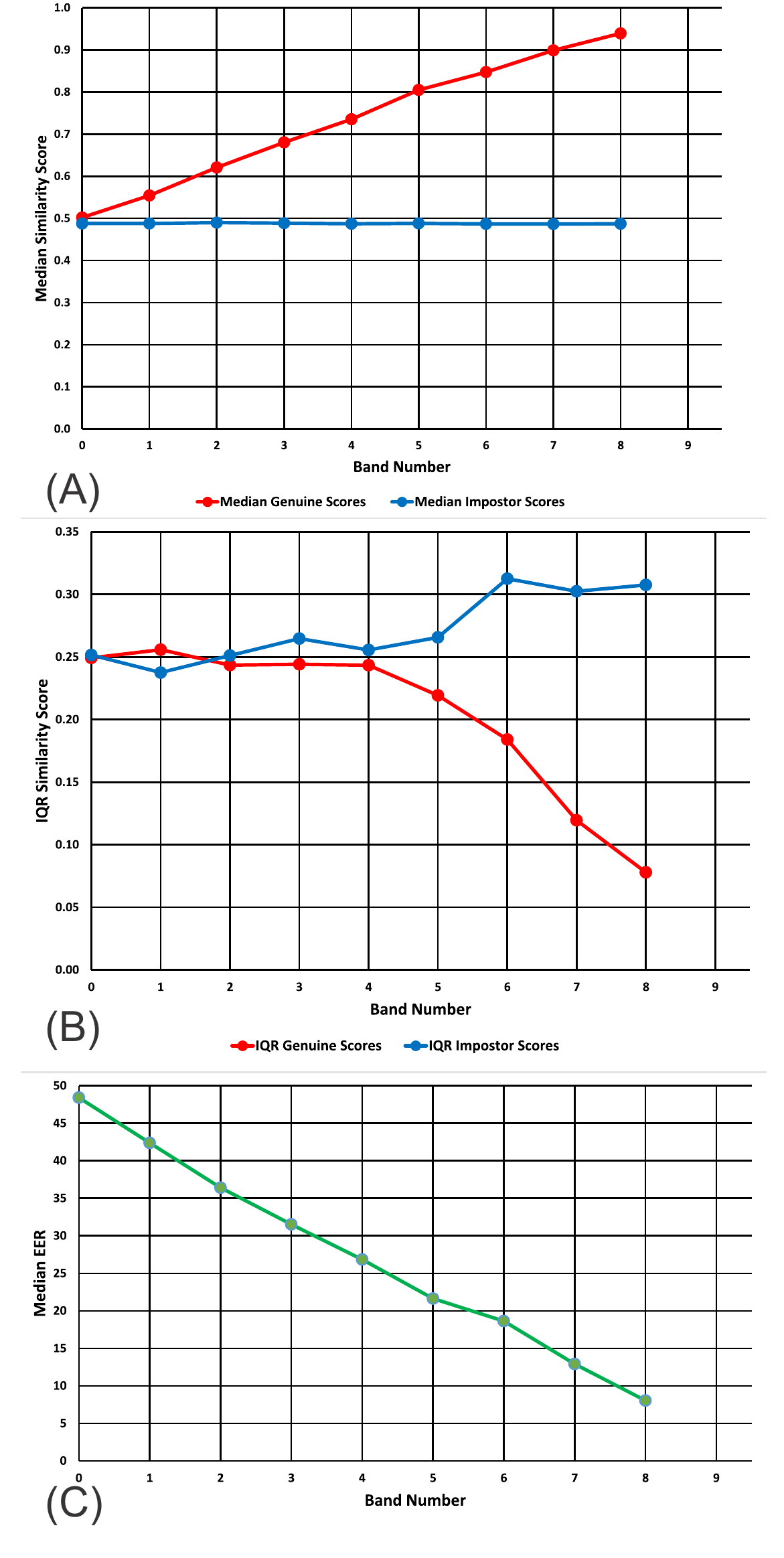}
    \caption{Medians and IQRs of genuine and impostor similarity scores across bands. (A)~Medians for SBA real dataset. (B)~IQRs for SBA real dataset. Ranges for the values in this table can be found in Appendix Tables 3, 4 and 5}
\end{figure}

\section{Evaluation of the Gait1 Dataset}
\label{sec:eval-gait1}

\subsection{Description of the Gait1 Dataset}
The Gait1~database is based on the Southampton Large Population Gait database of gait-related images, and videos~\cite{RN1337}. These databases are comprised of over 100~subjects tested on many sessions. Sessions can be as little as 1~minute apart. The analysis starts with a series of image frames while a subject walked. Binary silhouettes of the walking human were created~\cite{RN1601}. The silhouette extraction used chroma-key subtraction in conjunction with a connected components algorithm. The silhouettes were resized to 64~x~64~pixels. A series of image masks ($NMsk$~masks), like a mask for a horizontal band near the subject's waist, or the upper half of the silhouette, were applied to each subject silhouette, for each consecutive frame, and a time series for each mask type for each subject for each session was produced. At this stage, each subject was characterized by the time series of $NMsk$~masks. A cubic spline curve was fitted for the whole gait cycle, and 30~evenly spaced samples were taken from the whole curve, giving a single vector for each area mask used. These multiple vectors, for each 1~to~$NMsk$~mask, were reduced to a fewer dimensions, using canonical analysis. For each subject, the first feature was the first value in the final single vector, the second feature was the second value in this vector, and so on.
\subsection{Biometric Assessment of the Gait1 Dataset}
The ICCs for the Gait1~dataset were also available from~\citet{RN1513}. Once again, our goal was to divide up the data into ``temporal persistence'' bands with equal numbers of features in each band. We found that if we created band limits based on ICCs from 0.4~to~0.8 in steps of 0.1, we would have at least 7~features per band. There were simply too few features with lower ICCs to create reasonable bands. We randomly chose 5 of the 7 available features 10 times.  Standard ROC analyses were performed on each dataset and similarity score distribution characteristics (median and IQR) were saved and plotted as a function of band number.

\subsection{Results with Gait1 Dataset}
In Figure 5(A), we present the medians for the genuine and impostor distributions, for bands~4 (ICCs: 0.4~to~0.5) to 8 (ICCS: 0.8~to~0.9) for the real Gait1 dataset. In Figure 5(B), we present the interquartile ranges for the genuine and impostor distributions, for bands~4 to~8 for the real Gait1 dataset.

\begin{figure}
    \centering
    \includegraphics[width=0.65\textwidth]{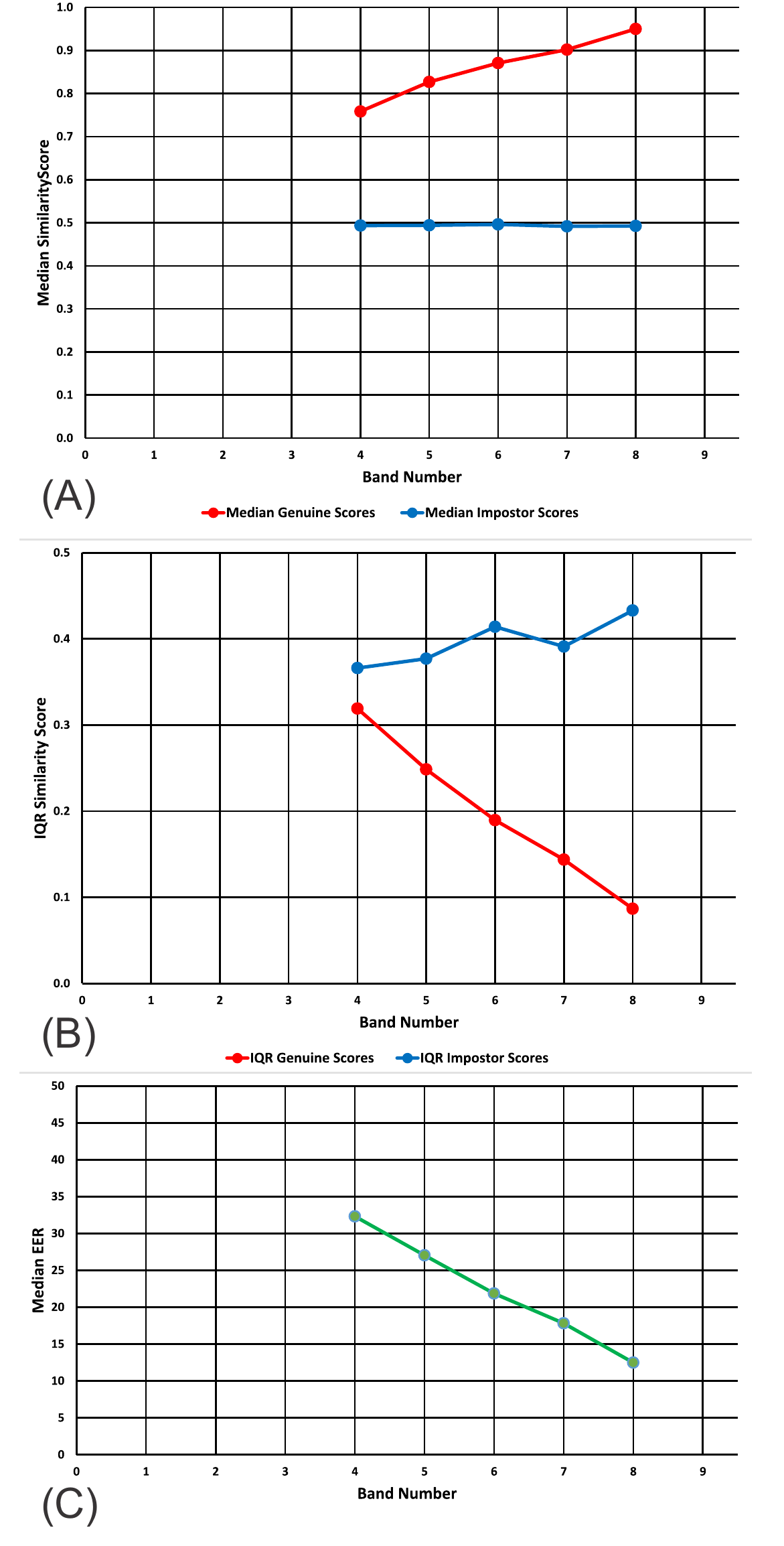}
    \caption{Medians and IQRs of genuine and impostor similarity scores across bands. (A)~Medians for Gait1 real dataset. (D)~IQRs for Gait1 real dataset. Ranges for the values in this table can be found in Appendix Tables 3, 6, and 7.}
\end{figure}

\section{Comparing Synthetic and Real Data Results}
\label{sec:compare}

\subsection{Discussion of Results}
For both the SBA and the Gait1 datasets, the pattern of results was similar to that obtained from the synthetic data. In all cases, the median of the distribution of genuine similarity scores increased linearly with ICC Band (Figures 3(A), 4(A) and 5(A), red lines). In all cases, the median of the distribution of imposter similarity scores was mostly flat and unrelated to ICC Band (Figures 3(A), 4(A) and 5(A), blue lines). In all cases, the IQR of the distribution of genuine similarity scores decreased with increasing ICC Band (Figures 3(B), 4(B) and 5(B), red lines). However, there was one obvious difference between the results for the datasets and the results for synthetic features. For our synthetic data, the IQR of the distribution of imposter scores was flat and unrelated to ICC band whereas for both real feature datasets, there was an increase in the IQR of the distribution of imposter scores as the ICCs increased (Figures 3(B), 4(B), and 5(B), blue lines).

To explain the observed differences in the behavior of the IQR of the distribution of imposter scores, we focus on one key difference between the synthetic data and the real data. The synthetic features were generally uncorrelated (up to simulation error), whereas within real datasets, the features would be intercorrelated. The degree of intercorrelation of real features increases with ICC, since high ICC features are less noisy than low ICC features. (The ICC is the ratio of subject variance to the sum of subject variance, session variance and error variance. As a general matter, high ICC features will have lower error variance than low ICC features.) Below, we report the results of a series of analyses investigating if this increasing intercorrelation could explain the differences noted between the biometric results using synthetic data and real data.

\subsection{Relationship between ICC Band and Median Intercorrelation between Features}
For both real datasets, the median intercorrelation (absolute values) increased with increasing ICC band (Figure 6). From this figure we can see that, for both datasets, there is a trend toward higher intercorrelation with increasing ICC band. It is not monotonic, but the increase is still obvious.

\begin{figure}
    \centering
    \includegraphics[width=0.95\linewidth]{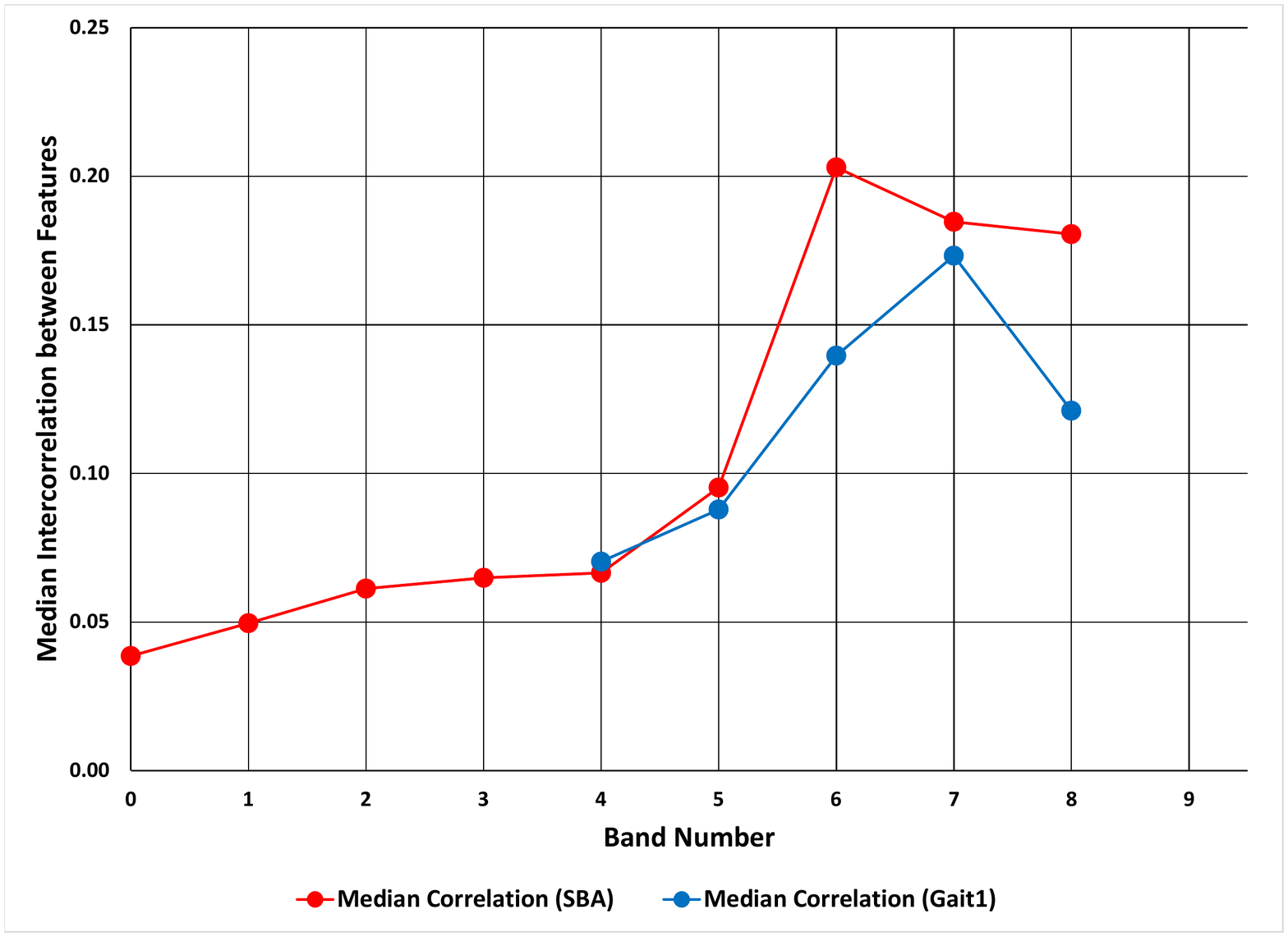}
    \caption{Relationship between ICC band and median intercorrelation between features. The results for the SBA dataset are presented in red and the results for the Gait1 dataset are presented in blue.}
\end{figure}

\subsection{Relationship between the IQR of Impostor Distributions and Median Intercorrelation}
To test if the IQR of the impostor distribution was related to the median intercorrelation between features in a dataset, we first combined the features from SBA Band~6 and Band~7. This yielded a combined dataset with 38~features. For 100~iterations, we randomly sampled, without replacement, 10~of the 38~features, computed the median Pearson r correlation coefficient (absolute value), and also computed the IQR of the impostor similarity scores for the random subset of features (Figure 7). From this figure we can see that the IQR of the impostor distribution increased linearly with increasing intercorrelation among features.

\subsection{The Effects of Decorrelation}
As a result of the analysis of the raw real data for the SBA dataset in Figure 4, we were interested in comparing the raw data with a decorrelated version of the data. If $N$ = number of subjects, $S$ = number of sessions and $K$ = number of features, then we collect the data in a matrix $X$ with $K$ columns and $r = N \cdot S$ rows. Let $Z_{r,r}$  be an $r$ by $r$ square matrix with every element 1. Define

\begin{figure}
    \centering
    \includegraphics[width=0.95\linewidth]{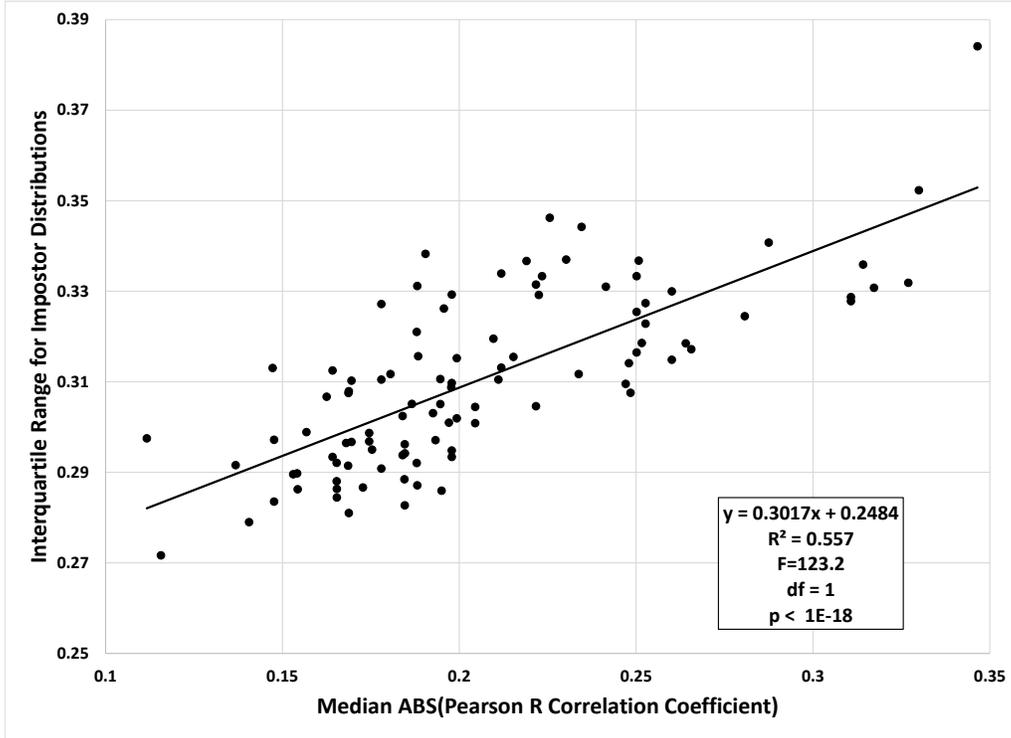}
    \caption{Relationship between IQR of the impostor distribution and median Pearson r correlation between features. As noted in text, the data are from the SBA real dataset.}
\end{figure}

\begin{equation}
    D = X - (\frac{1}{r}) Z_{r,r} X
\end{equation}
which is the data transformed so every random variable has zero mean. Then
\begin{equation}
    T = D(D^TD)^{-\frac{1}{2}},
\end{equation}
where the exponent of $-1/2$ represents the matrix square root of the inverse of a matrix, is a matrix of $K$ columns and $r$ rows, where all of the columns are completely uncorrelated. This is known as the inverse Cholesky factorization~\cite{RN1672}. See also: \color{blue}\href{https://blogs.sas.com/content/iml/2012/02/08/use-the-cholesky-transformation-to-correlate-and-uncorrelate-variables.html}{Decorrelation Webpage}.\color{black}

\begin{figure}
    \centering
    \includegraphics[trim={6em 2em 6em 2em}, clip, width=0.95\textwidth]{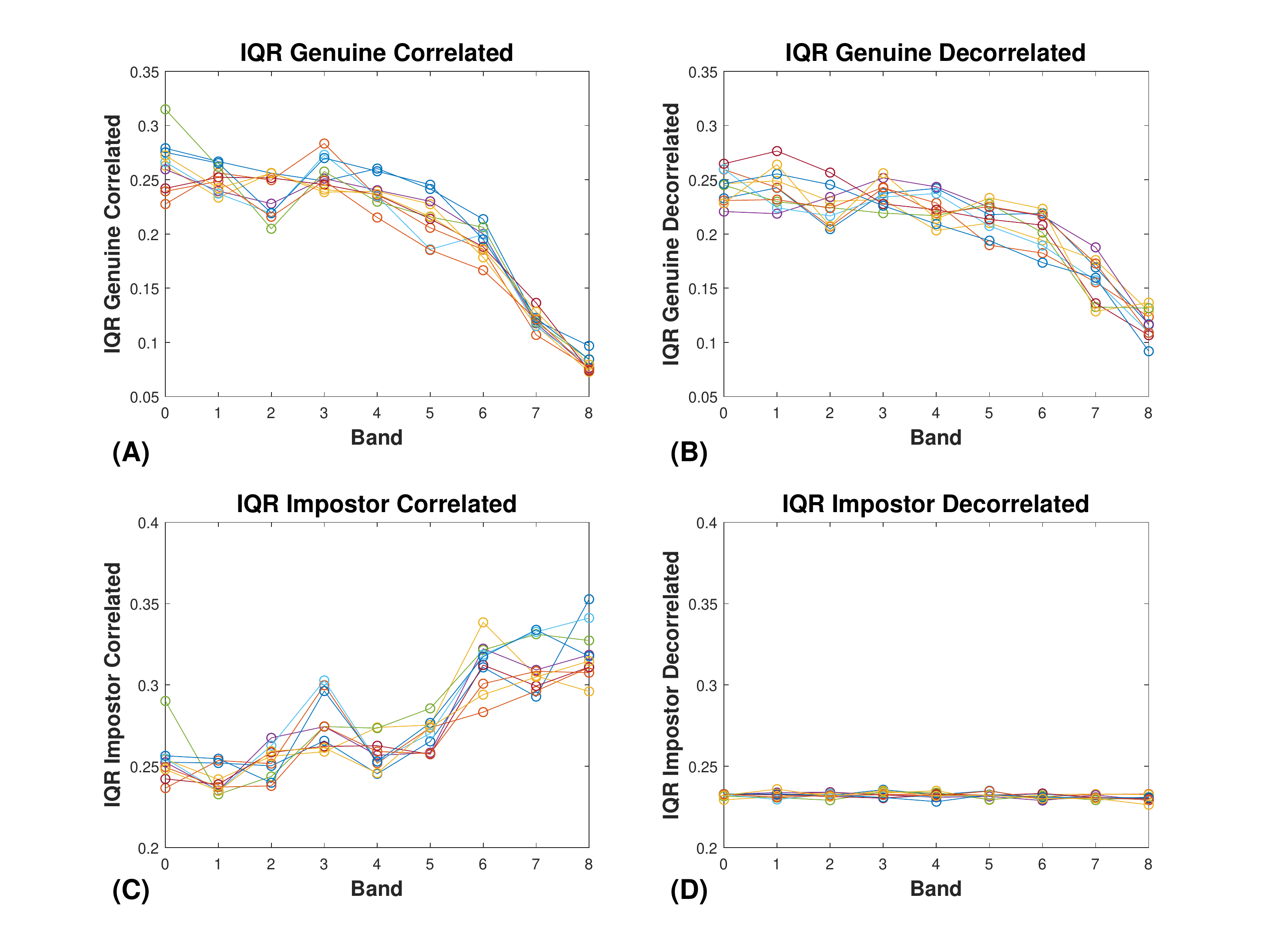}
    \caption{Effects of Decorrelation on IQRs. (A)~Relationship between the IQR of the genuine distribution and ICC Band for the SBA real dataset, prior to decorrelation. (B)~Relationship between the IQR of the genuine distribution and ICC Band for the SBA real dataset, after decorrelation. (C)~Relationship between the IQR of the impostor distribution and ICC Band for the SBA real dataset, prior to decorrelation. (D)~Relationship between the IQR of the impostor distribution and ICC Band for the SBA real dataset, after decorrelation.}
 \end{figure}

In Figure 8(A), we see the relationship between ICC band and the IQR of the genuine similarity scores for the original (intercorrelated) SBA dataset. For all the plots in Figure 8, there are 10~lines, each representing a different random selection of 10 of 19~features in each SBA band. In Figure 8(B), we see the same data as in Figure 8(A) after the features within each band have been decorrelated. In Figure 8(C), we present the relationship between ICC band and the IQR of the impostor distributions with the raw, intercorrelated data. In Figure 8(D) we see the same data as in Figure 8(C), after the features in each band have been decorrelated. The fact that removing the intercorrelations from the real data reproduces the findings of the synthetic data confirms that this is in fact the explanation for the observed difference.

\section{General Discussion}
\label{sec:discussion}
Increased temporal persistence improves biometric performance because of specific changes in the similarity score distributions for genuine and impostor samples. We show that, for both synthetic and real data, (1)~the median of the distribution of the genuine similarity scores increases with increasing ICC, (2)~the median of the distribution of the imposter similarity scores does not change as a function of ICC, and (3)~the IQR of the similarity scores for the genuine distribution declines with increasing ICC. This is consistent with the view that our synthetic features behave like our real data features and supports the use of these synthetic features to answer additional biometric-related questions.

The key difference between our real data features and our synthetic features has to do with the IQR of the impostor similarity scores, as ICC increases. With synthetic data, the IQR of the impostor distributions does not change with increasing ICC. (Recall that our synthetic features are all, at most, very weakly intercorrelated.)  However, with real data features, the IQR of the impostor distributions increase with increasing ICC. We have shown that this difference between the synthetic data and the real data is due to the fact that the degree of intercorrelation among real features increases with increased ICC. This is expected, because higher ICC features have less noise or error variance than lower ICC features. We have shown for real data that increased intercorrelation is associated with increased variance in these real datasets. Furthermore, we have shown that removing these intercorrelations with a decorrelation step removes the differences between the performance of real data and synthetic data.

Although our synthetic features revealed the same pattern as real data features in the main, there was a difference.  The difference stemmed from the fact that real features are intercorrelated and our synthetic features are uncorrelated.  This suggests the utility of developing synthetic features with intercorrelation patterns similar to real data.  The forward (as opposed to inverse) Cholesky transformation can create synthetic datasets given a variance-covariance matrix to emulate.  So, the inter-correlation pattern of any real dataset can be emulated in a synthetic dataset easily.  The question would remain, which real dataset, or how many different datasets, to emulate?  And what criteria to employ to make this decision? Also, it is not clear to us what the implications of doing this would be on the ICC of such a dataset.  This will require more thought and future study.

Interestingly, the relationship between increasing temporal persistence and characteristics of the genuine and impostor similarity scores was noted in our earlier publication~\cite{RN1513} (Figures~18 and~19 of that paper). In that paper, based on real biometric datasets, we found a statistically significant increase in the median genuine similarity score as the ICC of the data increased, whereas there was no statistically significant change in the median of the impostor similarity scores (Figure~18 in~\cite{RN1513}). We also found a statistically significant decrease in the IQR of the genuine similarity scores for datasets with higher ICC, whereas the change in the IQR for impostor distributions did not change in a statistically significant manner (Figure~19 of~\cite{RN1513}).  Our theoretical investigations with synthetic data have provided insight into why these phenomena occur and why temporal persistence is so valuable for biometric performance.

\section{Acknowledgement} 
We wish to thank Dr. Mark S. Nixon (Department of Electronics and Computer Science, University of Southampton, Southampton, United Kingdom) who made the gait data available to us.
We wish to thank, Dr. Raymond N. J. Veldhuis (Professor in Biometric Pattern recognition at The University of Twente, The Netherlands) who assisted us substantially in understanding and evaluating the "reliable-bits" literature (Section 2).
We wish to acknowledge the assistance of Dillon  J Lohr, a doctoral student in our group.  He was immensely helpful in preparing the \LaTeX\ version of the paper. 
The study was funded by 3 grants to Dr. Komogortsev:
(1)~National Science Foundation, CNS-1250718 and CNS-1714623, www.NSF.gov; 
(2)~National Institute of Standards and Technology, 60NANB15D325, www.NIST.gov; (3)~National Institute of Standards and Technology, 60NANB16D293. Dr. Stern's contributions were partially funded through Cooperative Agreement \#70NANB15H176 between NIST and Iowa State University, which includes activities carried out at Carnegie Mellon University, University of California Irvine, and University of Virginia.

\bibliographystyle{spbasic}
\bibliography{MyText}

\section{Appendix I}

\begin{verbatim}

***********************************************
R Code to Create Datasets of Synthetic Features
***********************************************

# Load relevant libraries
library(mise)
library(lme4)
library(matrixStats)
mise() # Clear out variables
# Close all previous connections
closeAllConnections() 
n<-1000 # Number of Subjects
k<-10 # Number of Features
s<-2   # Number of Sessions.  For Biometrics, s = 2.
ICC_Target<-0.7 # Target ICC
# Fill feature matrix with zeros
Feature_Matrix <- array(0,dim=c(n,k,s)) 
# Generate standard normal data for first session of
# all features
Feature_Matrix[,,1] <- rnorm(n*k,0,1)
# Set session 2 measurements equal to the first
# session measurements
Feature_Matrix[,,2] <- Feature_Matrix[,,1]
# At this point, Both session measurements match for
# each feature (for each individual) and ICC = 1.0.
# There is perfect # agreement.  We need to add 
# noise to features to obtain a lower ICC. The 
# noise we add will have a mean of 0 and a 
# SD = SD_Noise, calculated as:
SD_Noise<-sqrt((1-ICC_Target)/ICC_Target);
# Add this noise to all features
Feature_Matrix <- Feature_Matrix + rnorm(n*k*s,0,SD_Noise)
# Now we vertically concatenate the sessions
# to get a single matrix of dimension n*s x k
Both_Sessions <- rbind(Feature_Matrix[,,1],Feature_Matrix[,,2])
# Get the column Means
FeatureMeans<-colMeans(Both_Sessions) 
cat(sprintf('Column Means: 1:%0.3f 2:%0.3f 3:%0.3f 4:%0.3f 5:%0.3f 6:%0.3f
7:%0.3f 8:%0.3f 9:%0.3f 10:%0.3f\n',
FeatureMeans[1],FeatureMeans[2],FeatureMeans[3],FeatureMeans[4],
FeatureMeans[5],FeatureMeans[6],FeatureMeans[7],FeatureMeans[8],
FeatureMeans[9],FeatureMeans[10]))
FeatureSDs<-colSds(Both_Sessions)     
# Get the column SDs.
cat(sprintf('Column SDs  : 1:%0.3f 2:%0.3f 3:%0.3f 4:%0.3f 5:%0.3f 6:%0.3f
7:%0.3f 8:%0.3f 9:%0.3f 10:%0.3f\n',
FeatureSDs[1],FeatureSDs[2],FeatureSDs[3],FeatureSDs[4],
FeatureSDs[5],FeatureSDs[6],FeatureSDs[7],FeatureSDs[8],
FeatureSDs[9],FeatureSDs[10]))
# Create a Subject Number vector
Subject<-c(seq(1:n),seq(1,n))           
# Now we create a Session Vector
Session<-vector(mode='numeric',length = 0)
Session[1:n]<-1;
Session[(n+1):(n*2)]<-2;
# Horizontally concatenate the Subject Vector,
# the Session vector and the feature data into a
# single dataframe.
Feature_DF <-data.frame(Subject,Session,Both_Sessions)
names(Feature_DF)<-c("Subject","Session","Feat01","Feat02","Feat03",
"Feat04","Feat05","Feat06","Feat07","Feat08", "Feat09","Feat10")
cat(sprintf('\nWhich directory will the data be written to: %s\n\n',getwd()))
FeatureFileName<-paste0('SynthFeatSet_NSess_2_ICC_Targ_',
as.character(ICC_Target*10),'_NFeat_',as.character(k),'_NSubs_',
as.character(n),'.csv')
#Write out a .csv file with the synthetic features
write.csv(Feature_DF,FeatureFileName,row.names = FALSE) 
# Below we illustrate how to compute the ICC using
# variance components.
auto <- function(x) as.data.frame(VarCorr(lmer(x~(1|Feature_DF$Subject) +
(1|Feature_DF$Session) ,REML = TRUE, na.action=na.omit)))
lmer_results_apply<-apply(Feature_DF[,1:k+2], 2, auto)
ICCs<-vector(mode='numeric’, length=k)
for (i in 1:k){
  a<-unlist(lmer_results_apply[i])
  VarSub=as.numeric(a[10])
  VarSession=as.numeric(a[11])
  VarErr=as.numeric(a[12])
  TotalVariance<-sum(VarSub,VarSession,VarErr)
  ICCs[i]<-VarSub/TotalVariance
  cat(sprintf('For Feature %2.2d, VarSubject =   %0.3f, VarSession = %0.3f,
  VarErr = %0.3f, TotVar = %0.3f, ICC = %0.3f\n',
  i,VarSub,VarSession,VarErr,TotalVariance,ICCs[i]))
}
ICC_DF<-data.frame(ICCs)
names(ICC_DF)<-c("ICC")
ICCFileName<-paste0('SynthFeatSet_ICCs_NSess_2_ICC_Targ_',
as.character(ICC_Target*10),'_NFeat_',as.character(k),'_NSubs_',
as.character(n),'.csv')
# Write out .csv file with ICCs
write.csv(ICC_DF,ICCFileName,row.names = FALSE)
\end{verbatim}

\newpage
\section{Appendix II}
\begin{table}[ht]
    \centering
    \caption{Variability in estimates of median genuine and impostor similarity scores. Figure 3 (A)}
    \label{tab:median}
    \begin{tabular}{ccccccc}
        \toprule
        %Band & Median of Medians of Genuine Similarity Scores & Minimum of Medians of Genuine Similarity Scores & Maximum of Medians of Genuine Similarity Scores & Median of Medians of Impostor Similarity Scores & Minimum of Medians of Impostor Similarity Scores & Maximum of Medians of Impostor Similarity Scores \\
        {}     & Median of    & Minimum of   & Maximum of   & Median of    & Minimum of   & Maximum of   \\
        {}     & Medians of   & Medians of   & Medians of   & Medians of   & Medians of   & Medians of   \\
        {}     & Genuine      & Genuine      & Genuine      & Impostor     & Impostor     & Impostor     \\
        {}     & Similarity   & Similarity   & Similarity   & Similarity   & Similarity   & Similarity   \\
        Band   & Scores       & Scores       & Scores       & Scores       & Scores       & Scores       \\
        \midrule
        0	& 0.489	& 0.482	& 0.493	& 0.461	& 0.452	& 0.466 \\
        1	& 0.542	& 0.536	& 0.553	& 0.461	& 0.456	& 0.469 \\
        2	& 0.599	& 0.592	& 0.610	& 0.462	& 0.454	& 0.470 \\
        3	& 0.654	& 0.647	& 0.658	& 0.464	& 0.453	& 0.470 \\
        4	& 0.708	& 0.700	& 0.714	& 0.459	& 0.452	& 0.468 \\
        5	& 0.764	& 0.758	& 0.767	& 0.463	& 0.457	& 0.466 \\
        6	& 0.818	& 0.811	& 0.822	& 0.462	& 0.458	& 0.472 \\
        7	& 0.868	& 0.864	& 0.874	& 0.458	& 0.454	& 0.470 \\
        8	& 0.924	& 0.920	& 0.928	& 0.461	& 0.450	& 0.476 \\
        9	& 0.973	& 0.969	& 0.981	& 0.461	& 0.450	& 0.471 \\

        \bottomrule
    \end{tabular}
\end{table}

\begin{table}[ht]
    \centering
    \caption{Variability in estimates of IQR for genuine and impostor similarity scores. Figure 3 (B)}
    \label{tab:iqr}
    \begin{tabular}{ccccccc}
        \toprule
        {}     & Median of    & Minimum of   & Maximum of   & Median of    & Minimum of   & Maximum of   \\
        {}     & IQRs of      & IQRs of      & IQRs of      & IQRs of      & IQRs of      & IQRs of      \\
        {}     & Genuine      & Genuine      & Genuine      & Impostor     & Impostor     & Impostor     \\
        {}     & Similarity   & Similarity   & Similarity   & Similarity   & Similarity   & Similarity   \\
        Band   & Scores       & Scores       & Scores       & Scores       & Scores       & Scores       \\
        \midrule
        0	&	0.168	&	0.159	&	0.175	&	0.168	&	0.166	&	0.171		\\
        1	&	0.164	&	0.154	&	0.177	&	0.168	&	0.165	&	0.169		\\
        2	&	0.156	&	0.148	&	0.161	&	0.168	&	0.165	&	0.170		\\
        3	&	0.148	&	0.139	&	0.153	&	0.167	&	0.165	&	0.170		\\
        4	&	0.134	&	0.124	&	0.140	&	0.168	&	0.166	&	0.171		\\
        5	&	0.115	&	0.109	&	0.126	&	0.167	&	0.166	&	0.169		\\
        6	&	0.096	&	0.087	&	0.103	&	0.167	&	0.164	&	0.168		\\
        7	&	0.072	&	0.067	&	0.078	&	0.169	&	0.165	&	0.170		\\
        8	&	0.045	&	0.042	&	0.049	&	0.168	&	0.163	&	0.171		\\
        9	&	0.017	&	0.013	&	0.019	&	0.168	&	0.164	&	0.171		\\
        \bottomrule
    \end{tabular}
\end{table}

\begin{table}[ht]
\centering
\caption{Minumum and Maximum EER values for Figures 3, 4 and 5 (C)}
\begin{tabular}{ccccccc}
\toprule
 & Minimum   & Maximum   & Minimum & Maximum & Minimum & Maximum \\
Band & Synthetic & Synthetic & SBA     & SBA     & Gait1   & Gait \\    
\midrule
0 & 43.95 & 47.20  & 47.39  & 49.33 &  &  \\
1 & 36.10 & 38.08  & 41.84  & 43.59 &  &  \\
2 & 27.20 & 29.47  & 34.90  & 37.64 &  &  \\
3 & 20.40 & 21.98  & 30.18  & 31.96 &  &  \\
4 & 13.70 & 15.95  & 25.84  & 29.02 & 31.25   & 34.27   \\
5 & 8.00  & 10.00  & 20.47  & 22.82 & 25.00   & 28.57   \\
6 & 4.10  & 5.20   & 17.45  & 21.19 & 19.30   & 25.44   \\
7 & 1.50  & 2.15   & 10.61  & 16.14 & 15.56   & 18.96   \\
8 & 0.23  & 0.60   & 6.71   & 10.91 & 10.51   & 16.47   \\
9 & 0.00  & 0.01   &  &    &  & \\
\bottomrule
\end{tabular}
\end{table}

\begin{table}[ht]
    \centering
    \caption{Variability in estimates of median genuine and impostor similarity scores. Figure 4 (A)}
\begin{tabular}{ccccccc}
        \toprule
        {}     & Median of    & Minimum of   & Maximum of   & Median of    & Minimum of   & Maximum of   \\
        {}     & Medians of   & Medians of   & Medians of   & Medians of   & Medians of   & Medians of   \\
        {}     & Genuine      & Genuine      & Genuine      & Impostor     & Impostor     & Impostor     \\
        {}     & Similarity   & Similarity   & Similarity   & Similarity   & Similarity   & Similarity   \\
        Band   & Scores       & Scores       & Scores       & Scores       & Scores       & Scores       \\
        \midrule
0 & 0.502  & 0.491  & 0.515  & 0.488 & 0.482  & 0.490  \\
1 & 0.554  & 0.539  & 0.565  & 0.488 & 0.482  & 0.494  \\
2 & 0.621  & 0.607  & 0.641  & 0.490 & 0.486  & 0.494  \\
3 & 0.680  & 0.660  & 0.694  & 0.489 & 0.484  & 0.493  \\
4 & 0.735  & 0.706  & 0.752  & 0.487 & 0.484  & 0.492  \\
5 & 0.805  & 0.785  & 0.818  & 0.488 & 0.484  & 0.489  \\
6 & 0.847  & 0.840  & 0.855  & 0.487 & 0.485  & 0.491  \\
7 & 0.899  & 0.893  & 0.908  & 0.487 & 0.482  & 0.490  \\
8 & 0.939  & 0.934  & 0.943  & 0.487 & 0.478  & 0.491  \\ 
\bottomrule
\end{tabular}
\end{table}

\begin{table}[ht]
    \centering
    \caption{Variability in estimates of IQR for genuine and impostor similarity scores. Figure 4 (B)}
\begin{tabular}{ccccccc}
        \toprule
    {} & Median of    & Minimum of   & Maximum of   & Median of    & Minimum of   & Maximum of   \\
    {} & IQRs of      & IQRs of      & IQRs of      & IQRs of      & IQRs of      & IQRs of   \\
    {} & Genuine      & Genuine      & Genuine      & Impostor     & Impostor     & Impostor \\
    {} & Similarity   & Similarity   & Similarity   & Similarity   & Similarity   & Similarity   \\
    Band   & Scores   & Scores       & Scores       & Scores       & Scores       & Scores   \\
    \midrule
0 & 0.249 & 0.252  & 0.230  & 0.234  & 0.285  & 0.293  \\
1 & 0.256 & 0.238  & 0.220  & 0.233  & 0.275  & 0.257  \\
2 & 0.244 & 0.251  & 0.231  & 0.238  & 0.272  & 0.271  \\
3 & 0.244 & 0.265  & 0.212  & 0.243  & 0.261  & 0.293  \\
4 & 0.243 & 0.256  & 0.228  & 0.249  & 0.274  & 0.267  \\
5 & 0.219 & 0.266  & 0.189  & 0.253  & 0.233  & 0.282  \\
6 & 0.184 & 0.313  & 0.169  & 0.292  & 0.206  & 0.332  \\
7 & 0.120 & 0.302  & 0.104  & 0.293  & 0.147  & 0.357  \\
8 & 0.078 & 0.308  & 0.074  & 0.299  & 0.093  & 0.362 \\
\bottomrule
\end{tabular}
\end{table}

\begin{table}[ht]
    \centering
    \caption{Variability in estimates of median genuine and impostor similarity scores. Figure 5 (A)}
\begin{tabular}{ccccccc}
        \toprule
        {}     & Median of    & Minimum of   & Maximum of   & Median of    & Minimum of   & Maximum of   \\
        {}     & Medians of   & Medians of   & Medians of   & Medians of   & Medians of   & Medians of   \\
        {}     & Genuine      & Genuine      & Genuine      & Impostor     & Impostor     & Impostor     \\
        {}     & Similarity   & Similarity   & Similarity   & Similarity   & Similarity   & Similarity   \\
        Band   & Scores       & Scores       & Scores       & Scores       & Scores       & Scores       \\
        \midrule

4    & 0.758  & 0.739 & 0.769 & 0.493  & 0.492 & 0.497    \\
5    & 0.827  & 0.800 & 0.834 & 0.494  & 0.491 & 0.496    \\
6    & 0.871  & 0.852 & 0.892 & 0.496  & 0.493 & 0.499    \\
7    & 0.902  & 0.888 & 0.910 & 0.492  & 0.489 & 0.494    \\
8    & 0.950  & 0.941 & 0.952 & 0.493  & 0.491 & 0.494    \\
\bottomrule
\end{tabular}
\end{table}

\begin{table}[ht]
    \centering
    \caption{Variability in estimates of IQR for genuine and impostor similarity scores. Figure 5 (B)}
\begin{tabular}{ccccccc}
        \toprule
    {} & Median of  & Minimum of & Maximum of & Median of  & Minimum of & Maximum of   \\
    {} & IQRs of    & IQRs of    &IQRs of     & IQRs of    &  IQRs of   & IQRs of   \\
    {} & Genuine    & Genuine    & Genuine    & Impostor   & Impostor   & Impostor \\
    {} & Similarity & Similarity & Similarity & Similarity & Similarity & Similarity   \\
    Band  & Scores  & Scores     & Scores     & Scores     & Scores     & Scores   \\
\midrule
4 & 0.319  & 0.280  & 0.369  & 0.366  & 0.340  & 0.379  \\
5 & 0.249  & 0.234  & 0.277  & 0.377  & 0.348  & 0.407  \\
6 & 0.190  & 0.161  & 0.236  & 0.414  & 0.374  & 0.441  \\
7 & 0.144  & 0.119  & 0.164  & 0.391  & 0.367  & 0.403  \\
8 & 0.087  & 0.071  & 0.107  & 0.433  & 0.400  & 0.478  \\
\bottomrule
\end{tabular}
\end{table}
\end{document}